\documentclass[useAMS, usenatbib,usegraphicx]{mn2e}

\newcommand{\cmt}{cm$^{-3}$}

\newcommand{\kms}{km s$^{-1}$}

\newcommand {\apgt} {\ {\raise-.5ex\hbox{$\buildrel>\over\sim$}}\ }
\newcommand {\aplt} {\ {\raise-.5ex\hbox{$\buildrel<\over\sim$}}\ } 
\newcommand {\nh} {N$_2$H$^+$}
\newcommand {\hcn} {HCN}
\newcommand {\htcn} {H$^{13}$CN}
\newcommand {\hco} {HCO$^+$}
\newcommand {\co} {$^{12}$CO}
\newcommand {\tco} {$^{13}$CO}
\newcommand {\sig} {$\sigma$}
\newcommand {\M} {$M$}
\newcommand {\R} {$R$}
\newcommand {\sigR} {$\sigma - R$}
\newcommand {\MR} {$M - R$}
\newcommand {\chisq} {$\chi^{2}$}

\title[The linewidth-size relationship in the CMZ] {The linewidth-size relationship in the dense ISM of the Central Molecular Zone}
\author[R. Shetty et al.]{Rahul Shetty$^{1}$, Christopher N. Beaumont$^{2,3}$,  Michael G. Burton$^{4}$, \and Brandon C. Kelly$^{5}$, Ralf S. Klessen$^{1}$ \\
$^{1}$ Zentrum f\"ur Astronomie der Universit\"at Heidelberg, Institut f\"ur Theoretische Astrophysik, Albert-Ueberle-Str. 2, 69120 Heidelberg, Germany \\
$^{2}$ Institute for Astronomy, University of Hawai'i, 2680 Woodlawn Drive, Honolulu, HI 96822, USA \\
$^{3}$ Harvard-Smithsonian Center for Astrophysics, 60 Garden Street, Cambridge, MA 02138 \\
$^{4}$ School of Physics, University of New South Wales, NSW 2052, Australia \\
$^{5}$ Department of Physics, Broida Hall, University of California,
  Santa Barbara, CA 93106, USA}
\begin{document}

\date{Accepted 2012 June 25. Received 2012 June 25; in original form
  2012 April 25}

\pagerange{\pageref{firstpage}--\pageref{lastpage}} \pubyear{2010}

\maketitle

\label{firstpage}

\begin{abstract}

The linewidth (\sig) -- size (\R) relationship of the interstellar
medium (ISM) has been extensively measured and analysed, in both the
local ISM and in nearby normal galaxies.  Generally, a power-law
describes the relationship well with an index ranging from 0.2 -- 0.6,
and is now referred to as one of ``Larson's Relationships.''  The
nature of turbulence and star formation is considered to be intimately
related to these relationships, so evaluating the \sigR\ correlations
in various environments is important for developing a comprehensive
understanding of the ISM.  We measure the linewidth-size relationship
in the Central Molecular Zone (CMZ) of the Galactic Centre using
spectral line observations of the high density tracers \nh, \hcn,
\htcn, and \hco.  We construct dendrograms, which map the hierarchical
nature of the position-position-velocity (PPV) data, and compute the
linewidths and sizes of the dendrogram-defined structures.  The
dispersions range from $\sim$ 2 -- 30 \kms\ in structures spanning
sizes 2 -- 40 pc, respectively.  By performing Bayesian inference, we
show that a power-law with exponent 0.3 -- 1.1 can reasonably describe
the \sigR\ trend.  We demonstrate that the derived \sigR\ relationship
is independent of the locations in the PPV dataset where \sig\ and
\R\ are measured.  The uniformity in the \sigR\ relationship indicates
that turbulence in the CMZ is driven on the large scales beyond \apgt
30 pc.  We compare the CMZ \sigR\ relationship to that measured in the
Galactic molecular cloud Perseus.  The exponents between the two
systems are similar, suggestive of a connection between the turbulent
properties within a cloud to its ambient medium.  Yet, the velocity
dispersion in the CMZ is systematically higher, resulting in a scaling
coefficient that is approximately five times larger.  The systematic
enhancement of turbulent velocities may be due to the combined effects
of increased star formation activity, larger densities, and higher
pressures relative to the local ISM.


\end{abstract}

\begin{keywords}
ISM:\,clouds -- ISM:\,molecules -- ISM:\,structure -- stars:\,formation -- turbulence
\end{keywords}

\section{Introduction}\label{introsec}

Turbulence is observed on all scales larger than the size of the
densest star forming cores ($\sim$0.1 pc), and is considered a key
process influencing star formation in the interstellar medium
\citep[ISM, see][and references therein]{MacLow&Klessen04,
  McKee&Ostriker07}.  Atomic and molecular line observations provide
detailed information about the morphological, thermal, chemical, and
dynamical state of the ISM, such as the masses, velocities, and sizes
of a variety of structures, from the densest star-forming cores,
larger molecular clouds (MCs), and the surrounding, volume filling
diffuse gas \citep[e.g.][and references therein]{Young&Scoville91,
  Kalberla&Kerp09, Fukui&Kawamura10}.  Accurately measuring these
properties and the relationships between them is necessary for
developing a complete understanding of turbulence in the ISM,
including its role in the star formation process.

Spectral line observations of the ISM show strong power-law
correlations between the mass \M\ and linewidth \sig\ with their
projected size \R.  Observations of Milky Way MCs generally exhibit
exponents $\sim$2 and $\sim$ 0.2 -- 0.6 for the mass-size and
linewidth-size relationship, respectively \citep{Larson81, Solomon+87,
  Leisawitz90, Falgarone+92, Caselli&Myers95, Kauffmann+10a,
  Kauffmann+10b, Beaumont+12}.  Similar trends are observed for
extragalactic clouds \citep{Bolatto+08, Sheth+08}.  These results are
interpreted as evidence that MCs are virialized objects, and are
generally referred to as ``Larson's Relationships.''

That the \MR\ and \sigR\ relationships portray similar power law
scalings across a range in environments suggests that the underlying
physical processes driving ISM dynamics is ``universal''
\citep[e.g.][]{Heyer&Brunt04}.  However, it is not clear if this
``universality'' applies to systems significantly different than the
Milky Way or Local Group GMCs, as it has not been possible to
sufficiently resolve the ISM in external systems.  Measuring these
relationships in varying environments, such as (ultra) luminous
infrared galaxies ([U]LIRGs) or similar starbursting regions like the
Galactic Centre, is required to comprehensively understand the ISM
throughout the universe.


To understand the nature of turbulence in a starbursting environment,
we investigate the linewidth-size relationship of high density
structures in the Central Molecular Zone (CMZ) of the Galactic Centre
(GC).  The CMZ is denser than the ISM outside the GC, with densities
comparable to the star forming cores within MCs \citep[see][and
  references therein]{Morris&Serabyn96}.  The star formation rate in
the CMZ is $\sim$1.5 orders of magnitude higher than the rate measured
in the solar neighborhood \citep{Yusef-Zadeh+09}.  Therefore, the star
formation activity in the GC is in many ways similar to starbursting
systems like LIRGs.  Given its relatively close proximity ($\sim$ 8
kpc), the GC affords us the best opportunity to resolve the structure
of the ISM in starbursting environments, and probe the nature of the
dynamic ISM.

In this work, we measure the linewidth-size relationship in the CMZ
using Mopra observations of dense gas
tracers. \citet{Miyazaki&Tsuboi00} and \citet{Oka+98, Oka+01} present
an analysis of the Larson scaling relationships from Nobeyama
observations of CS and \co, respectively.  They show that the
\sigR\ relationship in the CMZ is similar to that in the Milky Way
disk, but with enhanced velocities.  We extend their analysis by
considering four high density tracers, \nh\ and \htcn\ with low
opacity, as well as \hcn\ and \hco\ with high opacity.  Further, the
Mopra coverage of the CMZ is fully (Nyquist) sampled, allowing for
thorough and systematic identification of dense clouds.  By
constructing hierarchical structure trees, or dendrograms
\citep{Rosolowsky+08}, we identify contiguous features in the observed
position-position-velocity (PPV) cubes of the CMZ.  One key difference
between our study and previous efforts is the manner in which
``clouds'' are identified.  Dendrograms map the full hierarchy of
structures, such that ISM structures which are not necessarily part of
the densest clouds are included in our analysis.  We assess the
linewidth-size relationship of the ensemble of these structures, and
compare it to the well known \sigR\ trends found in the local ISM.

This paper is organised as follows.  The next section describes the
observations of high density tracers in the CMZ.  In Section
\ref{methosec}, we describe the method we use to identify contiguous
structures in the observed data.  Section \ref{ressec} presents the
linewidth-size relationship.  In Section \ref{discsec} we discuss how
the measured \sigR\ relationship relates to the underlying turbulent
velocity field, and compare it to the trend in the Milky Way molecular
cloud Perseus.  We summarise our findings in Section \ref{summsec}.

\section{Observations and Analysis Technique}\label{obssec}

\subsection{Mopra survey of the CMZ}\label{mopobs}
In our analysis of the linewidth-size relationship of the CMZ, we use
the observations from the 22 metre Mopra millimetre-wave telescope,
which is operated by the Australia Telescope National Facility (ATNF).
The Mopra CMZ survey covers the 85-93 GHz window, covering 20 spectral
lines including a number of dense gas tracers like the \hcn, HNC and
\hco\ molecules, as well as the cold core species
\nh\ \citep{Jones+12}.\footnote{All data from the Mopra CMZ survey is
  publicly available at http://www.phys.unsw.edu.au/mopracmz.}  The
total area covered by the survey is 2.5$\times$0.5 deg$^2$, with
40\arcsec\ spatial resolution ($\sim$2 pc at a distance of 8 kpc), and
spectral resolution $<$ 0.9 \kms.  For our analysis, we use smoothed
datasets with 3.6 \kms\ spectral resolution of observations of the
lowest rotational transitions of four molecules: \nh\ (rest frequency
93.173 GHz), \hcn\ (88.632 GHz), \htcn\ (86.340 GHz), and
\hco\ (89.189 GHz).  Further information about the Mopra CMZ survey,
including reduction techniques, intensity maps, and velocity
structure, is described in \citet{Jones+12}.

\subsection{\tco\ FCRAO observations of the Perseus molecular cloud }\label{pobs}

In Section \ref{compsec}, we compare the linewidth-size relationships
in the CMZ to the Milky Way molecular cloud Perseus.  We use the
\tco\ J=1--0 line (110.201 GHz) data from the COMPLETE Survey
\citep{Ridge+06}.  The data are approximately Nyquist sampled, with a
pixel size of 23\arcsec\ (.028 pc at a distance of 250 pc), and have a
velocity resolution of 0.066 \kms.  \citet{Ridge+06} describes the
COMPLETE observations of Perseus, including \co\ and \tco, as well as
HI, dust extinction and emission maps of the cloud.

\subsection{Measuring the linewidth-size relationship}\label{methosec}

In order to measure the linewidth and sizes of structures, we first
need to identify the relevant structures using a well defined
criterion.  One rather simple definition of a structure is any
contiguous region with (approximately) constant density.
Unfortunately, we only have information about the intensity of a
number of spectral lines at a range of velocities, along lines of
sight (LoS) through the CMZ.  As the CMZ has extremely dense structure
throughout its volume, velocity integrated intensity maps would merge
a number of dense features along the LoS, thereby smoothing out the
intrinsic structure of the ISM in the resulting 2-dimensional (2D) map
\citep[e.g.][]{Pichardoetal00, Ballesteros-Paredes&MacLow02,
  Gammieetal03, Shetty+10}.  Accordingly, contiguous regions in the
3-dimensional (3D) PPV space should be able to more accurately
identify distinct structures.  As the ISM contains hierarchical
structure, direct clump decomposition methods may not accurately
identify different levels of substructure, especially in high density,
turbulent environments.  We thus employ the dendrogram decomposition
technique \citep{Rosolowsky+08} to identify and characterize
contiguous structures from the PPV cubes of the CMZ.

\begin{figure}
\includegraphics[width=90mm]{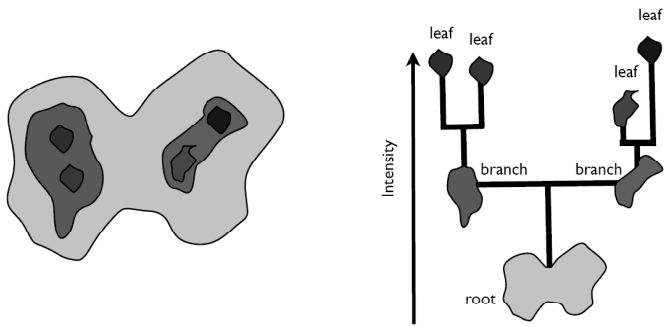}
\caption{Dendrogram of a hierarchical cloud.  Left: simple cloud with
  three levels of hierarchical substructure.  Right: Dendrogram, or
  structure tree, of the cloud.  Each distinct (closed contour)
  region, is identified as a structure (vertical lines).  Structures
  on the top of the tree are ``leaves.''  The structures which enclose
  multiple higher level contours are ``branches,'' and the structure
  which encloses all other structures is a ``root.''}
\label{dfig}
\end{figure}

We provide a brief overview of dendrograms here, and refer the reader
to \citet{Rosolowsky+08} for a more detailed description of the
structure decomposition algorithm.  Dendrograms provide a ``structure
tree'' which quantifies the hierarchical nature of any dataset.
Consider the cloud depicted in Figure \ref{dfig}.  The cloud is
composed of three different levels, or intensities, with the base
level (light blue structure) enclosing all higher levels.  Each
structure at a given level is distinguishable from adjacent structures
if a closed iso-intensity contour can enclose the whole structure. As
the intensities of the features do not merge in a contiguous manner,
they are identified as separate structures.  Both structures in the
second level encloses higher intensity regions, which are themselves
separated into two distinct structures.

The structure tree in Figure \ref{dfig} maps the hierarchical nature
of the model cloud.  Each vertical line corresponds to a distinct
structure in the cloud.  Those on the top of the tree are referred to
as ``leaves.''  The lowest level is the ``root'' of the tree.  Any
level which encloses multiple structures requires ``branches'' to
accommodate the hierarchy.  The leaves at top of the tree shows the
highest density structures which do not enclose any further
substructure.

The mapping illustrated in Figure \ref{dfig} can also be performed for
3D data, such as a PPV cube.  Each structure corresponds to an
iso-intensity region in PPV space.  Additional information can be
derived from the dendrogram-identified regions.  As the structures
from a PPV cube demarcate iso-intensity regions, the velocity
dispersions \sig\ of this structure may be directly computed.  We
compute the intensity weighted velocity dispersion of each structure
\begin{equation}
\sigma=\left [ \frac{\sum{I_v(v-\bar{v})^2}}{\sum{I_v}} \right]^\frac{1}{2}, 
\label{sig}
\end{equation}
where the summations are taken over all zones of the identified
structures, $I_v$ is the line intensity, $v$ is the LoS velocity, and
and $\bar{v}$ is the intensity weighted mean velocity,
$\sum(I_vv)/\sum I_v$.  We also define the size $R$ of each structure
using the projected area, $R = (N_{\rm proj} \Delta x \Delta
y/\pi)^\frac{1}{2}$, where $N_{\rm proj}$ is the number of projected
pixels, and $\Delta x \Delta y$ is the area of the resolution element.

To ensure that the dendrogram defined structures are well resolved, we
exclude any features with $R < 2$ pc, and $\sigma < 2$ \kms.  Since we
are interested in the dispersion of the contiguous and distinct
structures, and not the whole ISM in the CMZ, we also only consider
features with $R < 40$ pc.  Such a size limit may further minimise
contamination from the superposition of features along the LoS.
Finally, we also exclude any region which occurs within three voxels
(zones in a PPV dataset) from the boundary of the PPV cubes.  Under
these conditions, we minimise the likelihood of identifying features
dominated by observational noise, resulting in well defined and
contiguous structures PPV space.  We have adopted a uniform
``dendrogramming'' strategy such that each structure contains at least
50 voxels.  This condition ensures that the dendrogram identifies real
extended structures in the PPV cube, which are unlikely to be purely
noisy features.  From the measured value of \sig\ and \R, we can now
assess the linewidth-size relationship of the ensemble of dendrogram
defined structures.

\begin{figure}
\begin{center}
\includegraphics*[width=80mm]{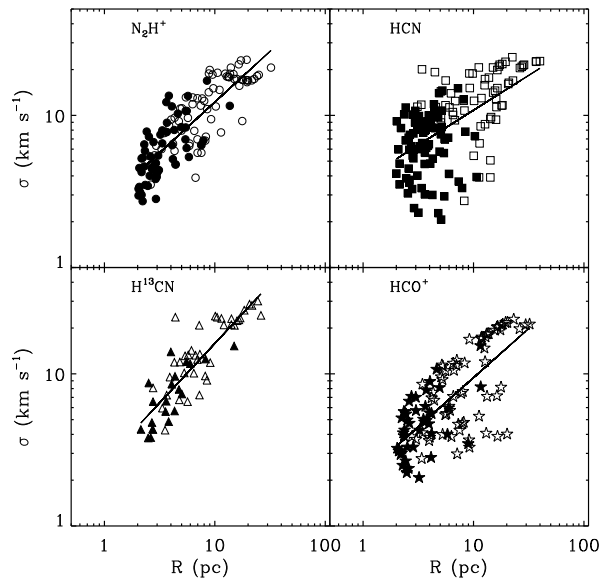}
\caption{Linewidth-size relationship in the CMZ, as measured within
  dendrogram identified structures in \nh, \hcn, \htcn, and \hco.
  Filled symbols correspond to ``leaves'' which do not enclose
  additional higher level structures.  Open symbols are structures
  which do contain higher level structures.  Lines show the best
  (\chisq) power-law fits.}
\label{4pans}
\end{center}
\end{figure}

\section{The linewidth-size relationship in the CMZ}\label{ressec}

Figure \ref{4pans} shows the \sigR\ relationship of the dendrogram
defined structures from the \nh, \hcn, \htcn, and \hco\ observations
of the CMZ. The largest structures identified in each tracer all have
dispersions of 20--30 \kms, which is equivalent to the global
linewidth of the CMZ.  These structures generally enclose higher
level, brighter features.  Towards smaller scales, the dispersions
systematically decrease, reaching $\sim$2 \kms\ in structures with
\R$\sim$2 pc.  The open symbols in Figure \ref{4pans} are regions
which contain at least one higher level structure.  The closed symbols
correspond to the leaves on the top of the structure tree (see Fig
\ref{dfig}).  The ensemble of structures exhibit a systematic decrease
in linewidths from the largest scale structures to the smallest,
brightest features in all the observed lines.

\begin{figure*}
\includegraphics[width=130mm]{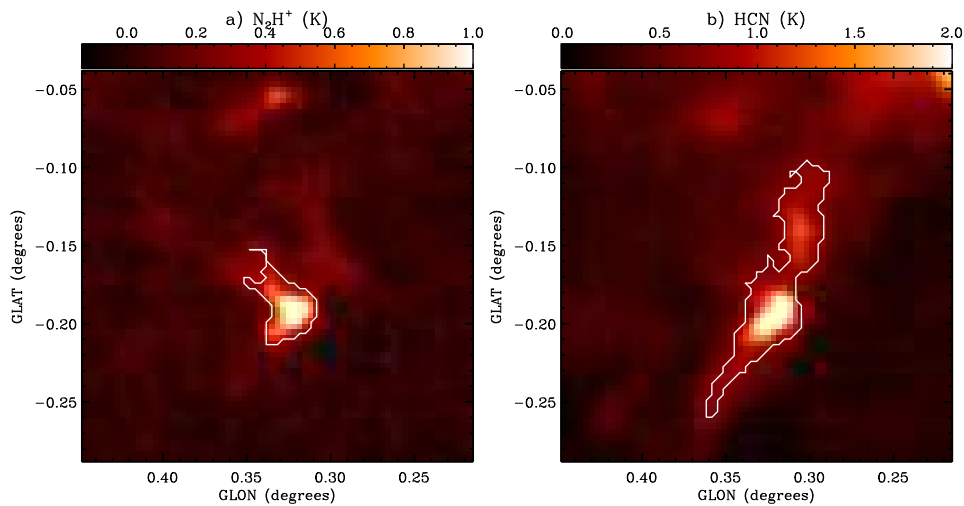}
\caption{A single a) \nh\ and b) \hcn\ channel map (corresponding to
  velocity $V_{\rm LSR} \sim$140 \kms) showing a dense filamentary
  region near to the Galactic Centre.  White contours show the
  cross-section of one dendrogram defined iso-intensity surface at
  this velocity.  The linewidths and sizes of the whole dendrogram
  structures are 2.8 \kms\ and 2.8 pc in \nh, and 2.1 \kms\ and 5.1 pc
  in \hcn, respectively.}
\label{slices}
\end{figure*}

That the structures in all tracers span a limited region in
\sigR\ space, as well as depict similar trends, provides confidence
that the apparent \sigR\ relationship in Figure \ref{4pans} is real.
\nh\ has a critical density of $\sim$$10^5$ \cmt, and is a good tracer
of cold, dense cores.  \hcn\ and \hco\ traces more extended dense gas
in the ISM (see Figure 2 in \citealt{Jones+12}), with similar critical
densities for each $J$=1-0 emission lines.  They can be moderately
optically thick in the CMZ, as analysed by \citet{Jones+12}.  Hence,
we perform the analysis on the weaker, but optically thin \htcn\ line,
obtaining essentially the same results; this indicates that optical
depth has not significantly biased the trends.  Both \hcn\ and
\nh\ exhibit hyperfine splitting, being split over $\sim$14 \kms\ into
3 groups of lines in the ratio 3:5:1.  Though this is less than the
typical turbulent width for the gas in the CMZ, multiple smaller scale
dendrogram structures in \nh\ and \hcn\ may actually be part of a
single feature.  On the other hand \hco, which is not split into any
sub-components, produces very similar results, indicating that the
derived linewidth-size relation of the ensemble is not significantly
affected by line splitting.

There are some differences between the linewidth-size trends from the
various tracers shown in Figure \ref{4pans}.  Most notably, the
\hcn\ and \hco\ have a number of larger scale structures with $R$
\apgt 3 pc and \sig \aplt 5 \kms\ that do not occur in \nh.  These
differences can be largely attributed to the relative extents of the
different tracers, and for some structures the foreground absorption
of \hcn\ and \hco\ emission.

For some of the large \R\ and low \sig\ points, a closer inspection of
these structures reveals that they are further divided into separate
features in the \nh\ dataset.  Figure \ref{slices} shows an example of
one such structure in \nh\ and \hcn.  The \hcn\ dendrogram-defined
contour corresponds to the leaf with the lowest \sig = 2.1 \kms, with
\R = 5.1 pc.  The dendrogram of the \nh\ feature, on the other hand,
is more compact with \R = 2.8 pc and \sig = 2.8 \kms.  \nh\ traces
cold, dense cores, and as it can be destroyed by reactions with CO, it
is expected to be present in the densest and most quiescent regions
where CO is depleted \citep[see][and references
  therein]{Bergin&Tafalla07}.  \nh\ is also found to be less extended
than \hcn\ and \hco\ \citep{Purcell+06, Purcell+09}. In addition, the
\hcn\ and \hco\ line emission has higher optical depth.  Hence the
dendrogram of an \hcn\ and \hco\ structure may therefore demarcate a
larger region surrounding a dense peak than does \nh, as shown in
Figure \ref{slices}.  Consequently, an extended \hcn\ structure may be
found to break up into a number of compact \nh\ features, with similar
measured linewidths.  The discrepancies in the \sig\ and \R\ provided
by the different tracers may not necessarily reflect morphological
variations of the gas, but rather diversity in chemical abundances.
The subdivision of features in the \nh\ observations results in a more
uniform \sigR\ relationship, and correspondingly a better defined
trend for a larger range in sizes.

As discussed by \citet{Jones+12}, there are strong foreground
absorption features in the \hcn\ and \hco\ datasets, possibly due to
the presence of the spiral arms along the LoS.  The sharp drop in
intensity of some of the large scale features across the velocity
channels corresponding to -52, -28, and -3 \kms\ leads the dendrogram
algorithm to identify multiple structures.  Consequently, such an
extended feature only spans a limited range in velocity, thereby
resulting in a structure with large \R\ but low \sig.  The division of
large scale structures due to foreground absorption along the LoS does
not occur as much in \nh\ and \htcn, as those molecules are not as
pervasive, resulting in substructure that can be more easily
differentiated from its background, as described above.

The \sigR\ trends shown in Figure \ref{4pans} indicate that there is a
large range of possible dispersions for a given size, reaching \apgt 7
\kms\ in the \hcn\ dataset.  Several factors can affect the amount of
substructure and measured properties from any observation, including
signal-to-noise ratio, resolution, chemical abundance inhomogeneities,
and radiative transfer effects (opacity, sensitivity to temperature
and density).  Further, local systematic effects may influence some
regions of the ISM more than others.  For instance,
\citet{Rodriguez-Fernandez&Combes08} suggest that infalling gas along
spiral arms and through the large cloud complex at longitude 1.3$^{\rm
  o}$ lead to the asymmetry of the CMZ.  We find that the measured
linewidths in this region are not particularly discrepant from the
ensemble, as the values of \sig\ and \R\ measured in that complex fall
within the range provided by the other structures.  The intrinsic
scatter of the measured \sig\ for structures with similar sizes may be
reflective of the differences in environmental conditions across the
ISM.
 
Despite the numerous physical processes that likely influence the gas
dynamics in the CMZ, Figure \ref{4pans} shows that the linewidths are
correlated with the sizes.  The Pearson's correlation coefficient of
the \nh, \hcn, \htcn, and \hco\ linwidth-size data (in log space) are
0.84, 0.58, 0.86, and 0.72, respectively.  The ensemble may thereby be
described as power-law relationships.  To quantify the power-law, we
fit
\begin{equation}
\left ( \frac{\sigma}{\rm km \, s^{-1}} \right ) =A \left(\frac{R}{\rm
pc} \right)^b
\label{pweqn}
\end{equation}
in log space to the dendrogram derived linewidths and sizes.  The fit
estimates the power law index $b$ and coefficient $A$.  We first
employ a \chisq\ fit, and in order to comprehensively account for the
uncertainties, we also perform a Bayesian linear regression analysis.

\subsection{Least squares (\chisq) fit}
As a first estimate of the power-law index and coefficient, we apply a
standard \chisq\ fit.  Each panel in Figure \ref{4pans} shows the
\chisq\ fit lines, and Table \ref{exptab} provides the best fit
parameters.  The maximum error in the best fit slopes is 0.06,
obtained from the \hco\ structures.  This error is the formal
\chisq\ error, and in this case is primarily driven by the large
number of points which results in rather low values.  The resulting
error estimates suggest that to a high degree of probability (3\sig )
the \nh\ and \hcn\ features have different slopes.  In principle, it
is possible that different molecules exhibit different \sigR\ slopes,
since they may span different regions in the ISM.  Such disparity may
be expected when considering tracers of different densities
(e.g. \co\ vs. \hcn), which would correspond to regions with variable
extents.  However, in this study we are focusing only on high density
tracers, and a direct comparison of the structures in the PPV
datacubes do not show obvious signs that the tracers span regions of
variable spatial extents.  Additionally, Figure \ref{4pans} suggests
significant overlap of \sig\ and \R\ for each of the tracers.
Therefore, the low error estimate in the \chisq\ fit may not
accurately characterize the uncertainty in the linewidth-size
relationship.  In order to rigorously account for the errors, as well
as ascertain the full range of possible best fit parameters, we employ
a Monte Carlo analysis and a Bayesian linear regression fit for
estimating the coefficient and power-law index.

\begin{table}
\centering
 \begin{minipage}{140mm}
  \caption{Best (\chisq) fit parameters of linewidth-size
    relationship$^1$}
  \begin{tabular}{cccccc}
  \hline
  \hline
  Tracer  & \nh & \hcn & \htcn & \hco  \\
 \hline
Power law index $b$  & 0.67 & 0.46 & 0.78 & 0.64 \\
Coefficient $A$  & 2.6 & 3.8 & 2.6 & 2.1 \\
\hline
\footnotetext[0]{$^1$ The formal 1$\sigma$ errors in $b$ and $A$ are
  all \aplt\ $0.06$ and 1.2, \\ respectively.}
\end{tabular}
\label{exptab}
\end{minipage}
\end{table}

\subsection{Uncertainties in the dendrograms}\label{uncert}

Dendrograms identify contiguous structures in the PPV cubes based on
the intensity at each position and LoS velocity.  Any uncertainties in
the intensities may lead to inaccurate identification of structures,
and thereby an erroneous \sigR\ relationship.  To assess the magnitude
of these uncertainties, we perform a simple bootstrapping analysis.

For this analysis we use the \nh\ dataset, which has a well defined
\sigR\ relationship.  We estimate a 1\sig\ uncertainty 0.04 K from a
region far from any significant emission.  For each realization in the
Monte Carlo simulation, we modify each voxel in the PPV cube by an
additive factor $\epsilon$, where $\epsilon$ is drawn from a normal
distribution with 0 mean and 0.04 K standard deviation.  After
constructing the structure tree, we fit the linewidths and sizes as
described above.

After performing this exercise 100 times, we measure a standard
deviation in the best fit parameters $b$ and $A$ to be \aplt 0.01.
This value is lower than the errors returned by the \chisq\ fit of the
original unperturbed dataset.  This test indicates that the variance
in the number of structures, as well as the uncertainty in the
estimated linewidths and sizes due to noise, hardly influences the fit
\sigR\ relationship.  We have performed a similar analysis on the
other three datasets, and as with the \nh\ analysis, we find that the
range in the best fit parameters to be smaller than the errors
computed in the \chisq\ fit.  We can thus be confident that the
dendrograms of the original dataset are not sensitive to noise in the
PPV datacubes.  We now turn our attention to the affect of resolution,
for which we employ a Bayesian linear regression analysis.

\subsection{Bayesian regression fit}

One of the primary advantages of Bayesian inference is that the
uncertainties in the measured data are rigorously and
self-consistently treated \citep[e.g.][]{Gelman+04, Kruschke11}.  A
Bayesian fit considers the error in each measured quantity, in our
case \sig\ and \R, to be drawn from some a priori defined
distribution.  The choice of the distribution should reflect the
uncertainties in the measurement.  Commonly, Bayesian inference
utilizes Markov Chain Monte Carlo (MCMC) routines to sample the
probability distribution of the fit parameters, given the measured
data and the assumed uncertainties.  The result of the Bayesian
inference is a joint probability $-$ or {\it posterior} $-$
distribution of the regression parameters.  This probability
distribution function (PDF) accounts for the assumed measurement error
distributions, and therefore provides well defined uncertainty
estimates on each of the fit parameters, as well as the correlations
between them.  We refer the reader to \citet{Kelly07} for details of
the Bayesian fitting algorithm we use in our analysis.\footnote{IDL
  routines for the Bayesian linear regression fitting algorithm is
  publicly available at http://idlastro.gsfc.nasa.gov.}

For our estimate on the uncertainties in the size we use the effective
spatial resolution of the observational datasets.  For the error
estimate in the velocity dispersion, we choose a conservative value of
2 \kms, which is about twice the resolution of the original
observations.  Note that though the resolution of the smoothed
datacube (3.6 \kms) is larger, any given contiguous region in a PPV
cube may produce a dispersion lower than the effective resolution, as
the dispersion is simply a measure of the deviation from the mean
value.  The smoothing primarily affects the very smallest features
with narrow lines\footnote{Line profiles along individual LoS are
  generally broader than 3.6 \kms}, which are excluded from our
analysis given our criteria described in Section 2.3.

In deriving the linewidths and sizes, we extract contiguous regions in
PPV space defined by the dendrogram algorithm.  Though this process
introduces additional sources of error, we have shown in Section
\ref{uncert} that these uncertainties are minimal, and hardly affect
the \sigR\ relationship.  Therefore, in our Bayesian fit we only
consider the uncertainties due to resolution (e.g. beam smearing).  We
are thus effectively assuming that resolution is the dominant source
of uncertainty, and that the errors in log(\sig) and log(\R) are
normally distributed, with dispersions 2 \kms\ and 2 pc, respectively.

\begin{figure}
\includegraphics[width=70mm]{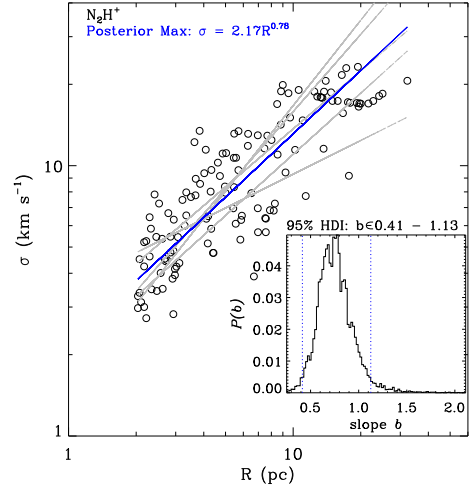}
\caption{Bayesian regression of linewidth-size relationship in
  \nh\ structures.  Open circles are the dendrogram derived linewidths
  and sizes.  The blue line corresponds to the fit with the highest
  probability.  The gray lines show five random draws from the
  Bayesian posterior.  The inset shows the marginal probability
  distribution $P(b)$ of the power-law index.  Vertical dashed lines
  indicate the 95\% highest density interval (HDI), which ranges from
  0.41 $-$ 1.13.}
\label{Bayescomp}
\end{figure}

Figure \ref{Bayescomp} shows the Bayesian fit of the
\nh\ \sigR\ relationship.  The peak of the posterior, corresponding to
$\sigma=2.2R^{0.78}$, is shown as the blue line.  As expected, this
line is similar to the \chisq\ fit.  The gray lines in Figure
\ref{Bayescomp} are five random draws from the Bayesian posterior.
The inset plot shows the marginal probability distribution of $b$.
The 95\% (2\sig) highest density interval, or HDI, is marked by the
vertical dashed lines, $b \in$ 0.41 $-$1.13.  This HDI provides a
quantitative prediction of the range in possible $b$, given the
uncertainties in the measurements of \sig\ and \R.  The distribution
of $b$ inferred by the Bayesian fit is six times larger than the
uncertainty in the \chisq\ estimate.

\begin{figure}
\includegraphics[width=70mm]{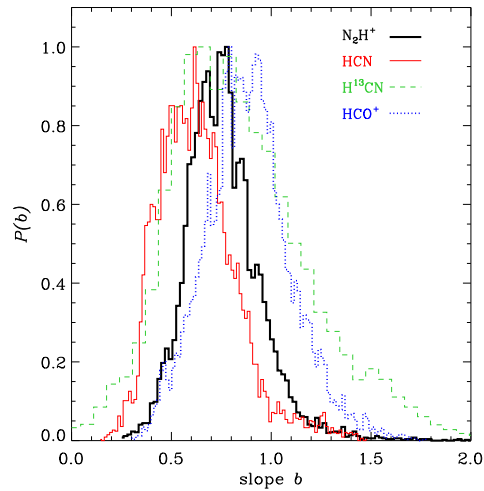}
\caption{Marginal distributions of the index $b$ from the Bayesian
  regression analysis of the \sigR\ relationship from \nh\ (thick
  black), \hcn\ (thin red), \htcn\ (dashed green), and \hco\ (dotted
  blue).  Table \ref{bexptab} provides the mode and the 95\% highest
  density intervals of these PDFs.}
\label{allhists}
\end{figure}

Figure \ref{allhists} provides the marginal PDFs of the fit indices
for all four tracers.  The \htcn\ PDF is noticeably wider, reflective
of the lower number of identified structures in that tracer
(Fig. \ref{4pans}).  Table \ref{bexptab} indicates the parameters from
the peak of the posterior and in the 95\% HDI for all the tracers.
There is substantial overlap between the PDFs from the different
tracers, in both the coefficient and the exponent.  It is common
practice in Bayesian inference to consider the full 95\% HDI as the
reasonable range of fit parameters, rather than any single
``best-fit'' point estimate, so that the measurement uncertainties are
taken into account.  Under this framework, we cannot conclude that the
evidence suggests a different \sigR\ relationship for the various
tracers.

\begin{table*}
 \centering
 \begin{minipage}{140mm}
  \caption{Results from Bayesian inference of linewidth-size
    parameters}
  \begin{tabular}{cccccc}
  \hline
  \hline
  Tracer  & \nh & \hcn & \htcn & \hco  \\
 \hline
Highest probability for $b$  & 0.78 & 0.62 & 0.66 & 0.79 \\
95\% HDI of $b$  & 0.41$-$1.13 & 0.27$-$1.02 & 0.21$-$1.66 & 0.45$-$1.30 \\
Highest probability for $A$  & 2.2 & 2.7 & 2.9 & 1.7 \\
95\% HDI of $A$  & 0.9$-$3.8 & 1.1$-$4.7 & 0.0$-$5.8 & 0.5$-$2.7 \\
\hline
\end{tabular}
\label{bexptab}
\end{minipage}
\end{table*}

\section{Discussion}\label{discsec}

\subsection{Do the growth of structures affect the \sigR\ relationship?}\label{growthsec}

The uniform \sigR\ trend over the full range in spatial scale shown in
Figure \ref{4pans} suggests some connection between the densest
structures and the larger scale ISM in the GC.  As we discuss further
in Section \ref{compsec}, this trend is also observed for clouds in
the Milky Way disk, and provides evidence for large-scale turbulent
driving.  One closely related question is whether the growth of dense
structures within a turbulent medium influences the underlying
velocity field.

To investigate the impact of dense clouds on the linewidth-size
relationship, we perform a similar analysis to that described in
Section \ref{methosec}, but with a few modifications.  First, we
assess the role of intensity weighting, simply by excluding the
weighting factor $I_v$ in Equation \ref{sig}.  Next, we also measure
the linewidth in random locations in the PPV cubes, using the
dendrogram defined contours placed at different locations throughout
the PPV cubes.  For the latter exercise, we shift the contours to a
different random location, such that the contours no longer demarcate
iso-intensity surfaces.  As a result, the shape of the contour is
preserved, but the PPV region within the shifted contour may consist
of a number of dense complexes, diffuse gas, or some combination of
both.  To ensure that results are statistically robust, we perform
numerous realizations of this exercise. Through these modified
analyses, we can investigate the extent which the growth of dense
structures and their internal processes influences the
\sigR\ relationship.

\begin{figure}
\includegraphics[width=70mm]{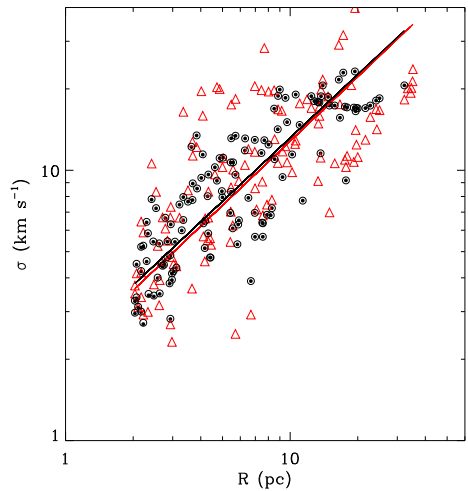}
\caption{Linewidth-size relationship of the \nh\ points.  Black open
  circles are the original \nh\ measurements, and the black line is
  the most probable Bayesian inferred fit $\sigma=2.2R^{0.78}$.  Black
  solid circles show the linewidths computed without any intensity
  weighting.  Red triangles are the measurements from random locations
  in the PPV cube, using dendrogram defined contours.  The red line is
  the most probable Bayesian inferred fit, $\sigma=2.1R^{0.79}$. }
\label{shiftpoints}
\end{figure}

Figure \ref{shiftpoints} shows the results from this modified
analysis.  The open circles are the original intensity weighted
\nh\ \sigR\ points, and the solid line shows the fit derived from the
maximum of the Bayesian posterior discussed previously and presented
in Figure \ref{Bayescomp}.  The smaller solid circles are the
measurements where there is no intensity weighting.  There is almost
no difference between these two sets of points, since the intensity
range within a given structure is limited.

The red open triangles in Figure \ref{shiftpoints} show the results of
one realization where the original dendrogram contours have been
randomly shifted to a different PPV location.  For this particular
realization, the contours have been shifted by a minimum of 50 voxels
in each of the three dimensions of the PPV cube, plus an additional
random value drawn from a uniform distribution between 0 and 20.
There is still a clear trend of increasing linewidth with increasing
size, with slightly more scatter than the original
\sigR\ distribution.  The solid red line shows $\sigma=2.1R^{0.79}$,
which corresponds to the maximum of the Bayesian posterior for this
dataset.  This result is very similar to the original \nh
relationship, and the HDIs for the two PDFs almost completely overlap.
We have performed ten different realizations of such a random shifting
of the dendrogram contours, varying the amount by which the contours
are shifted, and each one produces very similar results which are
statistically indistinguishable from the original \nh\ relationship.
We have also performed this analysis on the \hcn, \hco, and
\htcn\ datasets, and find the resulting \sigR\ relationships to be
statistically indistinguishable from the corresponding original
results.

As clouds begin to grow, self-gravity and other internal process will
begin to play a role in their subsequent dynamics.  If internal
processes dominate the ensuing dynamics of the clouds, we should find
a distinct difference between the \sigR\ computed from the dense
structures, and that measured from random regions irrespective of the
presence of dense structures.  That the measured \sigR\ relationship
can be reproduced regardless of the precise locations (in PPV space)
where the linewidths are measured indicates that the {\it dense cloud
  dynamics is predominantly governed by the underlying turbulent flow,
  with little contribution from internal sources of turbulence.}

In fact, a very similar analysis was carried out by \citet{Issa+90},
who concluded that the interpretation by \citet{Solomon+87} that GMCs
are bound arises due to the spatial and spectral crowding of clouds,
in combination with the method employed to identify clouds in PPV
cubes.  Our analysis is in agreement with \citet{Issa+90}.  The
measured \sigR\ relationship cannot fully reveal the dynamical state
of the dense clouds.  These results indicate that turbulence within
dense clouds are driven on the largest scales in the ISM, a topic we
further discuss in Section \ref{compsec}.

Besides revealing that turbulence must be driven on the largest
scales, the measured \sigR\ relationships may delineate a lower limit
to the turbulent driving scale.  Applying different statistical tools
on numerical simulations, such as structure functions and
$\Delta$-variance, \citet{Ossenkopf&MacLow02} showed that beyond the
driving scale \sig\ will remain nearly constant with increasing \R.
Moreover, dendrograms are capable of identifying the flattening of the
linewidth-size relationship towards the scales at which turbulence is
injected, as shown by \citet{Shetty+10} and \citet[][see their
Fig. 13]{Shetty+11b} using numerical simulations.  Although the \nh,
\hcn, and \hco\ \sigR\ pairs show some semblance of a decrease in $b$
at the largest scales (Fig. \ref{4pans}), there is no strong evidence
for such flattening.  Since \sig\ increases with \R\ up to the largest
scales in our analysis, we conclude that {\it turbulence in the CMZ is
predominantly driven on scales \apgt 30 pc}.  Note that though
turbulence is driven on the large scales, it produces a velocity field
which must be coherent on much smaller scales.\footnote{In the limit
of completely random and incoherent motions, there would be no
correlation between linewidth and size, or $b\approx0$, as discussed
by \citet[][see their Fig. 13]{Shetty+11b}.}  Quantifying this
coherence may further reveal additional properties of the turbulent
ISM.

\subsection{The relationship between the shape of dense structures and the \sigR\ relationship}\label{shapesec}

So far, we have defined the contiguous structures in PPV space via
dendrograms.  That is, their shapes are governed by the morphology and
velocity extent of the structures in the ISM.  In the analysis
described above, we have shown that the measured linewidths are rather
insensitive to the {\it location} in PPV space where they are
measured.  We now consider whether the precise {\it shape} of the
defined structures affect the linewidth measurement by analysing
regions with arbitrary shapes.  We consider two simple cases, one in
which the linewidths are measured in spherical regions, and one using
cubic regions from the PPV cube.

\begin{figure}
\includegraphics[width=70mm]{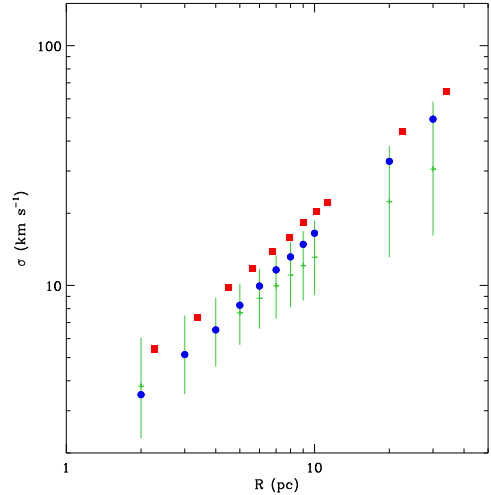}
\caption{Linewidth-size relationship of spherical (blue circles) and
  cubic (red squares) regions measured in the \nh\ PPV dataset.  For
  comparison, the horizontal green dashes are the predicted mean
  values of \sig\ from the Bayesian posterior (see
  Fig. \ref{Bayescomp}).  The green vertical lines show the 95\% HDI
  of the predicted values of \sig\ for each \R.}
\label{pp}
\end{figure}

Figure \ref{pp} shows the result of this exercise, comparing the
spherical and cubic derived linewidths to a representation of the
original dendrogram derived relationship from the \nh\ observations.
To portray the original measurement, we employ a ``posterior
prediction'' from the Bayesian regression of the \nh\ dataset shown in
Figure \ref{Bayescomp}.  Posterior prediction is a straightforward way
of comparing the Bayesian fit with new data.  In other words, it can
``predict'' new (or future) measurements based on the model derived
from the existing data \citep[see e.g.][]{Gelman+04, Kruschke11}.
From the Bayesian posterior, at each MCMC step we can draw a large
sample of \sig\ for chosen values of \R.  Due to the uncertainties and
the dispersion about the regression line, the posterior defines a
range in possible \sig\ values.  The 95\% HDI of the predicted \sig\
values from the dendrogram derived \sigR\ relationship is shown by the
green lines in Figure \ref{pp}, and the short horizontal lines show
the mean value.  Subsequent data that fall within the extent of the
vertical lines are considered to be consistent with the Bayesian
inferred \sigR\ relationship.

The blue circles in Figure \ref{pp} show the measured linewidths in
randomly extracted spherical regions in the \nh\ PPV cube.  The sizes
of the spheres are determined by the chosen value of the radius.
There are up to three spheres for each size considered, but as the
measured linewidths of each one are nearly identical, it is difficult
to distinguish the multiple structures.  The red squares are
measurements extracted from cubic regions.  The sizes \R\ of the
squares are computed in the same manner used to define the sizes of
the dendrogram structures, $R = \sqrt{{\rm A}/\pi} = L/\sqrt{\pi}$,
where $L$ is the length of one side of a face of the cube lying on the
2D plane spanning the spatial dimensions of the PPV cube.  The extent
of the spheres and cubes along the velocity dimension of the PPV cube
is rather arbitrarily chosen.  The measured velocity dispersion simply
scales with the velocity extent over which it is measured.  For the
cubes considered in Figure \ref{pp}, we simply set the extent in the
number of channels to be equal to the number of spatial pixels
defining the cube.  Similarly, for the sphere the number of spatial
pixels defining the ``radius'' sets the extent in velocity.  Had we
increased the extent in velocity, the cube would instead be a cuboid
(a 3D rectangle), and the sphere would be a cylindrical region
oriented parallel to the velocity axis, with semi-spherical ends.
Yet, the size would remain the same, due to its working definition.
Clearly, for the cubes and spheres in Figure \ref{pp} as we have
defined them the measured linewidths are dependent on the spectral
resolution of the dataset.

In general, the precise manner in which the sizes are defined will
affect any inferred \sigR\ relationship, as discussed previously by
\citet{Ballesteros-Paredes&MacLow02} and \citet{Shetty+10}.
Therefore, such caveats must not be ignored in interpreting any
apparent relationship.  Bearing these caveats in mind, Figure \ref{pp}
still reveals a few interesting properties.  

First, the \sigR\ relationship measured using spheres is similar to
that derived from dendrograms on the smaller scales where $R$\aplt 5
pc, but diverges at larger scales.  Given our definition of a sphere
and cube, along with our uncertainty estimates of the size (2 pc) and
dispersion (2 \kms), \sig\ scales (almost exactly) linearly with \R.
The \sigR\ relationship recovered from the spheres falls within the
95\% HDI of the Bayesian predicted \sig\ values.  Given the Mopra
resolution, it is therefore difficult to distinguish a
\sigR\ relationship obtained using the projected morphology of the
structures, where the most likely index $b<1$, from that derived
through randomly extracted spherical regions as we have defined it
here, which results in an index $b\approx$1.  Indeed, that a linear
relationship falls within the 95\% HDI of all the tracers (see
Tab. \ref{bexptab}) already alludes to this conclusion.  Within the
uncertainties, the indistinguishability between the \sigR\ trends
measured using the projected morphology of structures or using simple
spheres indicates that there is some class of shapes - which relates
the spatial and spectral extent of clouds - which would all recover
the underlying linewidth-size relationship.  This may be why previous
efforts using different structure identification schemes of GC clouds
have found similar results \citep[e.g.][]{Miyazaki&Tsuboi00, Oka+98,
  Oka+01}.

Second, the cubic \sigR\ relationship systematically recovers higher
linewidths than the spherical case, though the trend is still linear.
These linewidths occur beyond the 95\% HDI, suggesting that linewidth
measured in cubic region cannot reproduce the original dendrogram
derived \sigR\ relationship.  The offset by itself is not surprising,
since the cubic regions contain a larger number of zones at velocities
towards the edges of the defined region, thereby increasing the number
of velocities with large differences from the mean value,
correspondingly increasing the dispersion.  But, taken together with
the offset between the spherical and dendrogram defined regions, there
is an unambiguous decrease in linewidth from the cubes to the
dendrograms.  The variation may be reflective of the role of
turbulence, and the coherent velocities it generates, in sculpting the
dense structure in the ISM.

The shapes of structures formed in the ISM are likely to be governed
by the nature of the turbulence.  We have shown that it would be
difficult to distinguish between the \sigR\ relationship derived
considering spherical structures or a more standard approach using the
morphology of the structures.  But, as shown in Figure \ref{pp}, the
shape of the structures do matter, and arbitrary shapes with similar
sizes cannot generally reproduce the observed trend.  Detailed
analysis of the {\it shapes} of the structures, perhaps including more
robust definitions of the ``size,'' may shed light on this issue.  One
possible avenue is the use of a ``volume'' instead of size, for which
3D simulations of the ISM may aid in understanding how the shapes and
extents of structures in the PPV cube are correlated with the true
spatial morphology.

\subsection{Comparison with Milky Way GMCs}\label{persec}

Since the seminal work by \citet{Larson81}, the linewidth-size
relationship of the local ISM has been extensively measured.  It has
played a central role in theoretical efforts to explain the star
formation process (e.g. \citealt{Krumholz&McKee05,
Padoan&Nordlund11b}, see \citealt{MacLow&Klessen04, McKee&Ostriker07}
and references therein).  Given the uniform trends shown in Figure
\ref{4pans}, along with the results from local ISM studies, we are in
a position to perform a comparison between the two environments.

\begin{figure}
\includegraphics[width=70mm]{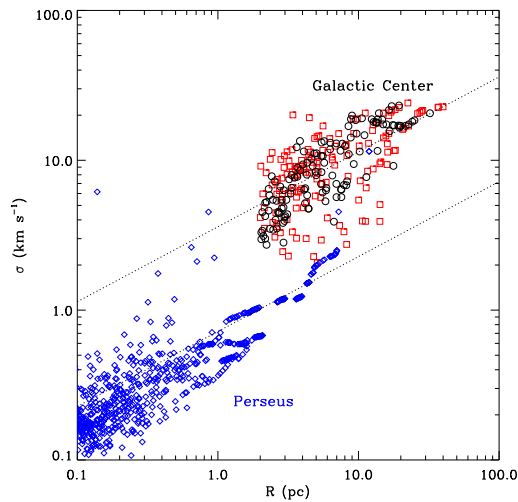}
\caption{Comparison of the linewidth-size of structures traced by
  \nh\ (black circles) and \hcn\ (red squares) in the Galactic Centre
  with \tco\ features from the Milky Way molecular cloud Perseus.
  Lower dashed line is the best fit relationship from
  \citet{Solomon+87}, $\sigma = 0.7R^{0.5}$.  Upper dashed line is the
  same relationship, but where the coefficient is 3.6.}
\label{MCcomp}
\end{figure}

As a first comparison, and to validate our measurement approach, we
compute the size-linewidth relationship from the dendrograms of
\tco\ observations of the Perseus molecular cloud.  Optically thin
\tco\ emission traces both the diffuse molecular gas, as well as dense
filaments and cores, within GMCs.  It is therefore analogous to the
dense gas tracers we consider for the CMZ, within which molecular gas
is pervasive.  Figure \ref{MCcomp} shows the \nh\ and
\hcn\ \sigR\ relationship in the CMZ, along with the trend measured in
Perseus.  The (\chisq) best fit line to the Perseus data is $\sigma =
0.62R^{0.54}$.  This relationship is similar to the best fit derived
by \citet{Solomon+87}\footnote{As indicated by \citet{Heyer+09}, the
  coefficient must be revised to 0.7 from the value reported by
  \citet{Solomon+87} due to improved estimates of the solar
  galactocentric radius.}, $\sigma = 0.7R^{0.5}$, and is shown as the
lower dashed line in Figure \ref{MCcomp}.

Clearly, the \sig\ -- \R\ trend in the GC is similar to the local ISM,
in that a comparable power-law index can account for the relationship
across a range in spatial scale \R\ $\in 2 - 40$ pc.  Yet, there is a
systematic enhancement in dispersions, resulting in a coefficient that
is about five times larger than the local ISM value (see Table
\ref{exptab}).  The upper dashed line in Figure \ref{MCcomp} shows the
local \sig -- \R\ relationship, enhanced by a factor of five.  There
is a reasonable agreement between this line and the linewidth-size
trend measured in the GC, in agreement with the analysis by
\citet{Oka+98, Oka+01} of individual and distinct CO clouds.

\subsection{Interpreting the linewidth-size relationship}\label{compsec}

The agreement between the \sig\ -- \R\ trend {\it within} a MC and
{\it between} individual MCs has been interpreted as evidence for the
``universality'' of turbulence by \citet{Heyer&Brunt04}.  Namely,
\citet{Heyer&Brunt04} suggest that turbulence observed in MCs
originates on the largest scale in the ISM \citep[see
  also][]{Ossenkopf+01,Ossenkopf&MacLow02,Brunt+09}.  In their
description, colliding flows in the diffuse ISM lead to the formation
of MCs.  While the MCs form at the interfaces of these colliding
flows, they inherit the turbulent characteristics of the ambient ISM,
leading to the observed ``universal'' \sig\ -- \R\ relationship;
smaller scale turbulence injection within MCs, e.g. due to internal
supernovae or stellar winds, appear not to contribute substantially to
the overall dispersions.

As we derive the linewidths in both the densest clouds and their
surrounding medium, we can assess the relationship of the clouds and
its environment.  The relatively uniform linewidth-size trend in the
CMZ, extending over an order of magnitude in spatial scales, may
indeed reflect a scenario in which the dense clouds inherit the
dynamic properties of the surrounding gas.  Further, our analysis of
the linewidths irrespective of the position within the PPV cubes (in
Section \ref{growthsec}) advocates for large scale driving as the
primary source of turbulence.  These results suggest that the transfer
of turbulent energy across spatial scales in the CMZ occurs in a
similar manner as in the more diffuse and quiescent ISM in the main
disc of the Milky Way.

Nevertheless, the larger coefficient in the \sigR\ relationship
attests to some fundamental difference in the CMZ and the local ISM.
The measured star formation rate in the CMZ is $\sim$50 times larger
than the rate in the solar neighborhood \citep{Yusef-Zadeh+09}.
Correspondingly, there is a higher rate of supernovae occurring in
this starbursting environment.  The variations in star formation
activity are tied to the differences between the environmental
properties of the main disk and CMZ.  Namely, the total ISM density,
vertical oscillation period \citep[see e.g.][]{Ostriker+10, Kim+11},
and perhaps even shear levels set by the rotational velocity at a
given Galactocentric radius, result in shorter dynamical times in the
CMZ, and the associated cloud formation and destruction timescales.
In the dense CMZ, the crowded sites of star formation and supernovae
result in a higher frequency of colliding streams, relative to the
main disc.  These distributed supernovae may be a primary source of
turbulence \citep{Ostriker&Shetty11, Shetty&Ostriker12}, which would
be external to any given dense cloud.  In the main disc, GMCs are
relatively isolated from each other, so that supernovae are not
distributed, resulting in a lower frequency of colliding shells.
Coupled with the larger densities and temperatures, the ambient
pressure in the CMZ is higher than the rest of the Milky Way.  The
enhanced turbulent velocity, and correspondingly the larger
coefficient in the linewidth-size relationship\footnote{See
  \citet{Heyer+09} for a discussion about the coefficient, and its
  relation to the dynamical state of the clouds.}, is likely
associated to the higher degree of star formation activity in the CMZ.

The relationship between turbulent velocities, cloud sizes, masses,
and the other physical properties of the ISM is germane for developing
theoretical descriptions of the star formation process.  Recent
theoretical efforts have focused on the role of turbulence in
regulating star formation \citep[e.g.][]{Krumholz+12, Ostriker+10,
  Ostriker&Shetty11}.  Numerical simulations of self-regulating
starbursting systems where distributed supernovae driven feedback
balances the vertical weight of the disk produce turbulent velocities
in dense gas of order 5 -- 10 \kms\ \citep{Shetty&Ostriker12}, similar
to the dispersion of the (presumably densest) clouds occurring on the
smallest scales in Figure \ref{4pans}.  In these models, the large
scale dynamics drives the evolution of gas on smaller scales,
including the velocity dispersion and ultimately star formation in
dense gas.  Quantifying the dynamical properties of the ISM will
further constrain the numerical simulations and the affiliated
description of star formation in similar starbursting environments.

As more observations resolve the structure and dynamics of a variety
of environments, a direct comparison of characteristics of the local
ISM will undoubtedly unfold.  Dust and/or spectral line observations
from lower density tracers will allow for estimates of the masses of
the structures in the CMZ \citep[e.g.][]{Longmore+12}.  Combined with
the \sigR\ relationship, the full dynamic state of the clouds and
ambient ISM should emerge.  {\it ALMA} will be ideal for carrying out
such an analysis, and reveal the similarities and differences between
starbursting regions like the CMZ and the more quiescent environments,
ultimately leading to a thorough understanding of the ISM.


\section{Summary}\label{summsec}

We have measured the linewidths and sizes of structures in the
observed \nh, \hcn, \htcn, and \hco\ PPV cubes of the CMZ.
Iso-intensity contours in the PPV cubes identify contiguous
structures, from which we measure the linewidths and projected sizes.
The ensemble of structures clearly portrays a linewidth \sig\ which
increases with size \R, with \sig\ $\sim$ 2 -- 30 \kms\ for structures
with \R\ $\in$ 2 -- 40 pc.

Employing a Bayesian regression analysis, we showed that the four
tracers all produce similar linewidth-size power-law relationships.
Fits from the tracers with the lowest noise levels, optically thin
\nh\ and optically thick \hcn, produce indices $b \in 0.3 - 1.1$, and
coefficients $A \in 0.9 - 4.7$ (Section \ref{ressec}).

By randomly shifting the iso-intensity contours within which \sig\ is
measured, we demonstrated that the measured \sigR\ trend is rather
independent of the presence of dense structures.  Therefore, turbulent
velocities are predominantly driven on the largest scales in the CMZ,
in agreement with interpretations of the ISM in the Milky Way disk.
That there is no strong evidence for a flattening of the
\sigR\ relationship in the CMZ suggests that turbulence is driven on
scales \apgt 30 pc (Section \ref{growthsec}).  We also discussed that
the morphology of structures and its velocities is (to some extent)
likely to be determined by the nature of the underlying turbulent but
coherent velocity field (Section \ref{shapesec}).

We compared the \sigR\ trend in the CMZ to that derived from Perseus.
A power law linewidth-size relationship with index 0.5, which
describes the local ISM well, reasonably captures the characteristics
derived in the CMZ.  This may be suggestive of the similarity in which
star forming clouds inherit the turbulent properties of the ambient
gas out of which they form.  However, the relationship in the CMZ
requires a factor of five enhancement in the coefficient relative to
the local \sigR\ power-law (Section \ref{persec}).  The starbursting
nature of the CMZ, along with its higher densities, temperatures, and
the associated pressure increase may be responsible for the larger
magnitudes in turbulent velocities.

\section*{Acknowledgements}

We are very grateful to Paul Jones for leading the effort in producing
the Mopra CMZ datasets.  We also thank Erik Bertram, Paul Clark,
Roland Crocker, Jonathan Foster, Simon Glover, Mark Heyer, David
Jones, Nicholas Kevlahan, Lukas Konstandin, Volker Ossenkopf, Eve
Ostriker, and Farhad Yusef-Zadeh for informative discussions regarding
the Galactic Centre, turbulence, and the linewidth-size relationship.
This paper improved from constructive comments by an anonymous
referee.  We acknowledge use of {\tt NEMO} software \citep{Teuben95}
to carry out parts of our analysis.  RS and RSK acknowledge support
from the Deutsche Forschungsgemeinschaft (DFG) via the SFB 881 (B1 and
B2) ``The Milky Way System,'' and the SPP (priority program) 1573,
``Physics of the ISM.''  MGB is supported by the Australian Research
Council for Discovery Project grant DP0879202.

\bibliography{citations}
\bibliographystyle{mn2e}

\label{lastpage}
\end{document}